\documentclass[ %
 reprint,
superscriptaddress,
nobibnotes,
amsmath,amssymb,
aps,
maxbibnames=3
pra,
prb,
rmp,
prstab,
prstper,
floatfix,
]{revtex4-2}
\usepackage{graphicx}
\usepackage{dcolumn}
\usepackage{bm}
\usepackage[hidelinks]{hyperref}
\hypersetup{colorlinks=true,linkcolor=blue}
\usepackage[mathlines]{lineno}
\usepackage{xcolor, soul} 
\usepackage{xurl}

\usepackage{academicons}
\usepackage{orcidlink}
\usepackage{tabularx}
\usepackage{adjustbox}
\usepackage{multirow}
\usepackage{caption}
\usepackage{subcaption}
\usepackage{enumitem}
\usepackage{physics}
\hypersetup{breaklinks=true}

\begin{document}

\preprint{APS/123-QED}

\title[Exploring the test of the no-hair theorem with LISA]{Exploring the test of the no-hair theorem with LISA}

\author{Chantal Pitte \orcidlink{https://orcid.org/0009-0009-0524-7292}}
    \email{chantal.pitte@cea.fr}
\affiliation{IRFU, CEA, Universit\'{e} Paris-Saclay, F-91191, Gif-sur-Yvette, France.}

\author{Quentin Baghi \orcidlink{https://orcid.org/0000-0002-1555-9283}}
\affiliation{Astroparticule et Cosmologie, Universit{\'e} de Paris, CNRS, Paris, 75013, France}
\affiliation{IRFU, CEA, Universit\'{e} Paris-Saclay, F-91191, Gif-sur-Yvette, France.}

\author{Marc Besançon \orcidlink{https://orcid.org/0000-0003-3278-3671}}
\affiliation{IRFU, CEA, Universit\'{e} Paris-Saclay, F-91191, Gif-sur-Yvette, France.}

\author{Antoine Petiteau \orcidlink{https://orcid.org/0000-0002-7371-9695}}
\affiliation{IRFU, CEA, Universit\'{e} Paris-Saclay, F-91191, Gif-sur-Yvette, France.}
\affiliation{Astroparticule et Cosmologie, Universit{\'e} de Paris, CNRS, Paris, 75013, France}

\date{\today}

\begin{abstract}

In this study, we explore the possibility of testing the \textit{no-hair} theorem with gravitational waves from massive black hole binaries in the Laser Interferometer Space Antenna (LISA) frequency band. Based on its sensitivity, we consider LISA's ability to detect possible deviations from general relativity (GR) in the ringdown. Two approaches are considered: an agnostic quasi-normal mode (QNM) analysis and a method explicitly targeting the deviations from GR for given QNMs. Both approaches allow us to find fractional deviations from GR as estimated parameters or by comparing the mass and spin estimated from different QNMs. However, the estimated deviations may vary depending on whether we rely on prior knowledge of the source parameters from a pre-merger or inspiral-merger-ringdown analysis. Under some assumptions, the second approach targeting fractional deviations from GR allows us to recover the injected values with high accuracy and precision. We obtain $5\%, 10\%$ uncertainty on $\delta \omega, \delta \tau$ respectively, for the $(3,3,0)$ mode, and $3\%, 17\%$ for the $(4,4,0)$ mode. As each approach constrains different features, we conclude combining both methods would be necessary to perform a better test. In this analysis, we also forecast the precision of the estimated deviation parameters for sources throughout the mass and distance ranges observable by LISA. 

\end{abstract}


\maketitle

\section{\label{sec1}Introduction}

The first detection of gravitational waves (GWs) by the LIGO-Virgo-KAGRA (LVK) Collaboration \cite{Abbott_2009, ligo_ad_2015, virgo, kagra}, produced by the coalescence of the black hole binary (BHB) GW150914 \cite{Abbott_2016}, marked the beginning of the GW astronomy era. At the same time, its detection opened a window to probe physics beyond the standard model and general relativity (GR). Since that first detection, the scientific community has been eager to test GR in the strong field regime \cite{ligo_testGR1, ligo_testGR_150914, ligo_testGR2, ligo_testGR3}. 

Indeed, GR can be tested, for instance, through the \textit{no-hair} theorem by analyzing the \textit{ringdown} phase. The no-hair theorem is a prediction of GR that states that any Kerr black hole (BH) is characterized by only two parameters: its mass and its spin ($M_f, a_f$)~\cite{Israel, Carter, Robinson, Hawking_72}. In the last stage of the coalescence, i.e., the ringdown, the perturbed remnant BH will settle down through the emission of GWs. In perturbation theory (PT), the gravitational radiation of a Kerr BH can be written as a superposition of damped sinusoids~\cite{1973ApJ_T, 1973ApJ_P, Vishveshwara_1970, Detweiler_1977, chandra_75, 1980ApJ_239_292D}. The GW strain in the plus and cross polarizations reads

\begin{align}\label{spheroidal}
    h_{+}(t) - i h_{\times}(t) =  \sum_{lmn} h_{lmn}(t){}_{-2}S_{lmn}(a_f\Tilde{\omega}_{lmn}; \theta, \varphi),
\end{align}
where 

\begin{equation}\label{h_spheroidal}
    h_{lmn}(t) =  A_{lmn} e^{i(\phi_{lmn} - \Tilde{\omega}_{lmn}t)}.
\end{equation}

The equations describing a perturbed Kerr BH were first derived by Teukolsky in 1973~\cite{1973ApJ_T}. In collaboration with Press~\cite{1973ApJ_P}, he described the subsequent radiation of GWs, decomposing the solutions in terms of spin-weighted spheroidal harmonics ${}_{s}S_{lmn}(a_f\Tilde{\omega}_{lmn};\theta, \varphi)$, with $s=-2$ denoting the spin of the field. These spheroidal harmonics depend on the eigenvalues $\tilde {\omega}_{lmn}$, the spin, the phase $\varphi$ and the angle $\theta$ between the normal of the orbital plane and the observer. Each harmonic is then labeled by $(l,m,n)$ as polar and azimuthal angular numbers and overtone, respectively. These numbers are integers taking values: $ |s| \leq l, |m| \leq l, 0 \leq n$, where the (2,2,0) is, in general, the dominant carrier of energy. The eigenvalues of the \textit{Teukolsky master equation} are complex frequencies $\tilde {\omega}_{lmn}$ known as quasi-normal modes (QNMs)~\cite{chandra_75, Detweiler_1977}. The real part of these solutions denotes the oscillation frequency, while the imaginary part corresponds to the inverse of the damping time:

\begin{equation}
    \Tilde{\omega}_{lmn} = \omega_{lmn} + i/\tau_{lmn}.
\end{equation}
The QNMs' values are characterized by the metric structure; thus, for a remnant Kerr BH, they depend only on the mass and spin~\cite{1973ApJ_T, 1973ApJ_P, Echevarria, Finn_1992}. In contrast, the amplitude and phase associated with each mode $(A_{lmn}, \phi_{lmn})$ are related to their excitation in the pre-merger phase, thus depending on the initial BHs' parameters \cite{Kamaretsos_2012, Kamaretsos_2012_2}.

One approach to probe the no-hair theorem is to use \textit{BH spectroscopy}~\cite{chandra_75, Detweiler_1977, 1980ApJ_239_292D, Kokkotas_1999, Nollert_1999}. In GR, the values of the BH spectrum, i.e. the QNMs, are defined solely by the mass and spin of the final BH. Therefore, when studying the spectrum of the remnant BH, one can recover those two parameters with a parametrization~\cite{Echevarria, Finn_1992, Dreyer_2004, Berti_2006}. Then, two pairs of mass and spin obtained from different QNMs should be consistent with each other. In an alternative theory to GR, the values of the complex frequencies might deviate from those of GR \cite{Molina_2010, tattersall_2018, arun2022}. Thus, the pairs of mass and spin derived from each QNM in an alternative theory, might not longer be consistent with each other~\cite{Kamaretsos_2012, Gossan_2012}. Note that more than one QNM is required to perform this analysis. There also exists another method where only one QNM is needed; it involves the comparison of the estimated parameters of mass and spin in the pre-merger phase with the estimated parameters of mass and spin from the QNM~\cite{ligo_testGR_150914, ligo_testGR1, ligo_testGR2, ligo_testGR3, Isi_2019}.

To this day, while the fourth observational run (O4) is ongoing, almost a hundred sources have been confidently observed by the LVK Collaboration~\cite{catalogue3}, while another hundred sources are additionally potential candidates~\cite{events_alert}. In several events, the characteristic ringdown waveform has been observed~\cite{ligo_testGR3}. Moreover, small hints of spherical harmonics beyond the dominant ($2,2$) mode have been found for some events in \cite{gennari2023}, using the \textit{tidal effective one-body post-merger} (TEOBPM)~\cite{Damour_teob} waveform model. However, the signature of QNMs seems to hide below the noise floor, despite the possible presence of the first overtone of the dominant QNM, i.e. $(2,2,1)$, which has been inferred with low significance for the first event GW150914 \cite{Isi_2019, Giesler_2019} (see the discussion on its detectability \cite{cotesta_22, isi2022revisiting, Isi_2023, Carullo_2023}, and further analyses \cite{Ma_2023, Crisostomi_2023, wang2023frequencydomainperspectivegw150914ringdown, correia2024lowevidenceringdownovertone}). As the sensitivity of current and future interferometers increases \cite{ligo_ad_2015, ET, LISA_Proposal2017}, we expect to detect more QNMs, hopefully already from the O4 analysis. However, the question of whether we can unmistakably observe a deviation from GR remains. Up to the O3 catalog, various analyses have been made with results always in agreement with GR \cite{ligo_testGR_150914, ligo_testGR1, ligo_testGR2, ligo_testGR3, Ghosh_2021, maggio2023tests}. More reliable analyses to confidently discriminate an alternative theory would rely on a null hypothesis comparison in a Bayesian approach. However, this endeavor presents quite a challenge since it would require that BH's spectra should be solved for alternative theories. Various developments in computing beyond-GR BH's spectra have been undertaken, primarily for static or slowly rotating BHs in different alternative theories, e.g.~\cite{Molina_2010, tattersall_2018, Bl_zquez_Salcedo_2016, Pierini_2021}, see nonetheless \cite{cano2023} for rapidly rotating BHs in effective field theories and \cite{metrics3, metrics4} for a new numerical approach to compute the QNM values in beyond-GR theories. In the lack of waveform catalogs including deviations from GR, the best we can do is to allow for deviations of GR in a model-independent way, both in the injection and search templates.

As the Laser Interferometer Space Antenna (LISA) passed the adoption phase, the moment to detect massive black hole binary (MBHB) mergers with high signal-to-noise ratio (SNR) approaches~\cite{LISA_Proposal2017, LISA_redbook}. With high SNR, high precision is also expected. Therefore, in sources where the detectability of higher harmonics is possible~\cite{Pitte}, we also expect to detect various QNMs. 

In this exploratory study, we address the question of the extent to which LISA becomes distinctively sensitive to a deviation from GR in the ringdown phase of BHB coalescence. In a similar context, possible deviations from GR with LISA sources have been studied in~\cite{toubiana2024} using the pSEOBNRv5HM waveform~\cite{Brito_2018,maggio2023tests}. In that work, the full inspiral-merger-ringdown (IMR) was used to find deviations present only in the ringdown. While using the full IMR is an advantage for low SNR sources, for high SNR, systematic errors in the full IMR waveform might bias the estimated parameters of the remnant if eccentricity, precession, or higher harmonics are not accounted for. 
In this work, we consider a more flexible prior knowledge of parameters, assuming a raw posterior distribution of the final BH parameters estimated from the IMR as uniform priors for this analysis. We use two approaches to assess LISA sensitivity to detect deviations from GR, namely an \textit{agnostic} approach and a \textit{deviations} approach targeting the deviations in specific QNMs. We will also discuss the outcome of different assumptions on the priors for the first case.

The paper is divided as follows. In Sec.~\ref{sec2}, we describe the response of LISA to GWs. Sec.~\ref{sec3} is dedicated to the methodology implemented, including a description of the data, the templates, and the likelihood computation. In Sec.~\ref{sec4}, we show and comment on the results. We then forecast how the GR test precision relates to SNR in Sec.~\ref{sec_GR_SNR}. We conclude in Sec.~\ref{sec6}.

\section{\label{sec2}LISA response to waveforms}

LISA consists of three spacecrafts (S/C) in heliocentric orbits and arranged in a triangular formation with arms of $2.5 \times 10^{6}$ km. One reason for this particular setup is that with different combinations of individual phasemeter measurements, one can construct multiple synthetic interferometers~\cite{Vallisneri_2005}. 

GWs crossing the beam paths will imprint a frequency shift between the emitted and received light at the detectors. The measurement at the end of each arm is called the link response. Each link response is defined as 
\begingroup\makeatletter\def\f@size{9}\check@mathfonts
\def\maketag@@@#1{\hbox{\m@th\normalsize#1}}%
\begin{align}\label{eq:response1}
\begin{split}
    y_{rs}(t_r) \simeq & \frac{1}{2 \left( 1 - \textbf{k} \cdot \hat{\textbf{n}}_{rs} (t_r)\right)} \left[ 
    H_{rs}\left( t_r - L_{rs}(t_r) - \right. \right. \\  
    & \left. \left. \textbf{k} \cdot \textbf{x}_r(t_r)\right)  - H_{rs} \left( t_r - \textbf{k} \cdot \textbf{x}_s(t_r) \right) \right].
\end{split}
\end{align}
\endgroup
We use geometric units $c=G=1$ throughout the whole study. Lower indices $r$ and $s$ take values from 1 to 3 for the three S/C, representing the light-receiving and the light-sending spacecraft, respectively. The S/C positions are defined by \textbf{x}$_{r,s}$.  $L_{rs}$ is the arm's length between the two S/C. The vector $\textbf{k}$ defines the direction of propagation of the GW, while $\hat{\textbf{n}}_{rs}$ denotes the direction of the beam. Lastly, $H_{rs}$ is the source's gravitational strain projected on the arm. It reads
\begin{align}
\begin{split}
    H_{rs}(t) & = \, \left( h_{+} (t) \cos 2\psi \, - \right.  \\
    & \left. \hspace{8mm} h_{\times}(t) \sin 2\psi\right) \, \hat{\textbf{n}}_{rs}(t) \cdot \textbf{e}_{+} \cdot \hat{\textbf{n}}_{rs}(t) \\
    & + \, \left( h_{+}(t) \sin 2\psi \, + \right.  \\
    & \left. \hspace{8mm} h_{\times}(t) \cos 2\psi\right) \, \hat{\textbf{n}}_{rs}(t) \cdot\textbf{e}_{\times} \cdot  \hat{\textbf{n}}_{rs}(t),
\end{split}\label{eq:response2}
\end{align}
where $\psi$ is the polarization angle, defined as the rotation angle along the line of sight between the source frame and the observational frame, set to zero in the following. The polarization tensors, \textbf{e}$_{+,\times}$, are defined in the traceless-transverse gauge 
\begin{equation}
    h^{TT} = \textbf{e}_+ h_{+} + \textbf{e}_\times h_{\times},
\end{equation}
as
\begin{subequations}\label{eq_tensor}
\begin{align}
    \textbf{e}_{+} = & \textbf{u} \otimes \textbf{u} - \textbf{v} \otimes \textbf{v}, \\
    \textbf{e}_{\times} = & \textbf{u} \otimes \textbf{v}  + \textbf{v} \otimes \textbf{u}.
\end{align}
\end{subequations}
Vectors \textbf{v} and \textbf{u} together with the propagation vector \textbf{k} in spherical coordinates locate the source in the observational frame,
\begin{subequations}\label{eq_vector}
\begin{align}
    \textbf{u} = & \{\sin \lambda, \cos \lambda, 0 \},\\
    \textbf{v} = & \{-\sin \beta \cos \lambda, -\sin \beta \sin \lambda, \cos \beta \},\\
    \textbf{k} = & \{-\cos \beta \cos \lambda, -\cos \beta \sin \lambda, -\sin \beta \},
\end{align}
\end{subequations}
with $\beta, \lambda$ the ecliptic latitude and longitude respectively. 

LISA does not work like a regular Michelson interferometer, as its interferometry is synthetically performed on the ground in the post-processing. Given that LISA arms will not have an equal length in space, nor would they be stationary, particular linear combinations with time delays of links are needed to cancel the noise produced from fluctuations in the laser \cite{Tinto_1999, Tinto_2004, Armstrong_1999, Tinto_2002, Estabrook_2000, Vallisneri_2005} among others noises. The combinations that allow for the laser noise suppression are known as time delay interferometry (TDI) channels ($X$, $Y$, $Z$) and will be produced off-line, such that 
\begin{align}
        X  & =  \, \left( 1 - D_{121} - D_{12131} + D_{1312121} \right) \left(y_{13} + D_{13} y_{31}\right) \nonumber \\
     & \, - \left(1 - D_{131} - D_{13121} + D_{1213131} \right) \left(y_{12} + D_{12} y_{21}\right),
\end{align}
where $D_{ij}$ is the delay operator \cite{Bayle_2021}, 
\begin{equation}\label{delay}
    D_{rs} \, f(t) = f\left(t - L_{rs}(t)\right),
\end{equation}
with $f(t)$ any function dependent on t. The combination of operators is written as
\begin{equation}
     D_{i_1 i_2 \cdots i_n} = D_{i_1 i_2} D_{i_2 i_3} \cdots D_{i_{n-1} i_n}.
\end{equation}
One should perform a cyclic permutation of the S/C indices to obtain channels $Y$ and $Z$. The S/C positions are required to compute the light travel time (LTT) of the armlength $L_{rs}(t)$ between them. For this reason, we use the orbits computed for each S/C from ESA LISA science orbit files \cite{ESA_orbits} generated with the \texttt{LISA orbits} package \cite{bayle_2022_6412992}. The LTT will affect the delay operators Eq.~\eqref{delay} as well as the projection of the strain onto the links in Eq.~\eqref{eq:response1}. 

To perform data analysis, optimal combinations of the channels $X$, $Y$, and $Z$ can be found to obtain quasi-orthogonal channels. They are defined as \cite{Tinto_1999, Tinto_2002, Tinto_2004, Estabrook_2000, Armstrong_1999, Prince_2002}
\begin{subequations}
    \begin{align}
        A = &\frac{1}{\sqrt{2}}(Z - X), \\
        E = &\frac{1}{\sqrt{6}}(X - 2Y + Z), \\
        T = &\frac{1}{\sqrt{3}} (X + Y + Z).
    \end{align}
\end{subequations}

In this analysis, we work only with channels $A$ and $E$, as $T$ is almost blind to GWs in the low-frequency regime~\cite{Prince_2002}. Even though that channel could provide complementary information about the source, the computational effort is unrewarding in this case.

\section{Methodology}\label{sec3}

We aim to test for the presence of deviations from GR in the ringdown of MBHBs with a LISA prototype pipeline. To this end, we developed a code~\cite{lisaring} capable of generating ringdown waveforms with the response of LISA. Anticipating the detailed description that will be given in a separate paper~\cite{lisaring}, we describe the main features of the code in the following section.

\subsection{Data}\label{sec_data}

The analysis procedure consists of generating a toy model describing the ringdown phase of an MBHB as the sum of damped oscillations with the response of LISA. The sum of damped oscillations in the ringdown of MBHB is given by
Eq.~\eqref{spheroidal}, with 

\begin{equation}\label{strain}
    h_{lmn}(t) = \frac{M_f}{D_l} A_{lmn}(\Xi,t)e^{i(\phi_{lmn}(\Xi, t) -  \tilde{\omega}'_{lmn}t)}.
\end{equation}
The complex frequency with a tilde includes an allowed fractional deviation from GR in the real and the imaginary part, as first introduced by~\cite{Kamaretsos_2012}:
\begin{subequations}\label{GR+dev}
    \begin{align}
        \omega'_{lmn} &= \omega_{lmn}^{\mathrm{GR}}(M_f, a_f) (1 + \delta \omega_{lmn}), \\ 
        \tau'_{lmn} &= \tau_{lmn}^{\mathrm{GR}}(M_f, a_f) (1 + \delta \tau_{lmn}).
    \end{align}
\end{subequations}
Then,
\begin{equation}\label{dev}
    \Tilde{\omega}'_{lmn} = \omega'_{lmn}+ i/\tau'_{lmn}.
\end{equation}
The $\mathrm{GR}$ index indicates the values obtained within the GR framework. There are different ways to compute complex frequencies; we recommend~\cite{Berti_2009} and~\cite{franchini2023testinggeneralrelativityblack} for reviews on this topic. Here, we make use of the \texttt{qnm} package \cite{Stein}, which is based on a spectral eigenvalue approach \cite{Cook_2014}.

Furthermore, the letter $\Xi$ in the amplitude and phase stands for the intrinsic redshifted mass and spin parameters of the progenitors $\Xi = (m_1, m_2, \chi_1, \chi_2)$, and $D_l$ is the luminosity distance of the source. The spheroidal harmonics in Eq.~\eqref{spheroidal} can be decomposed as~\cite{1973ApJ_P}
\begin{align}\label{spho_sphe}
    {}_{s}S_{lmn} = {}_{s}Y_{lm} \, + \sum_{l \neq l'} \frac{\bra{sl'm} \mathfrak{h}_1 \ket{slm}}{l(l+1) - l'(l'+1)} + \cdots ,
\end{align}
where we drop the dependence on $(a_f\Tilde{\omega}_{lmn};\theta ,\varphi)$ for clarity, and where~\cite{1973ApJ_P}
\begin{equation}\label{clebsh}
    \mathfrak{h}_1 = a_f^2\omega^2 \cos^2{\theta} - 2a_f \omega s \cos{\theta},
\end{equation}
and
\begin{equation}
    \bra{sl'm} \mathfrak{h}_1 \ket{slm} = \int_{\Omega} d\Omega \, {}_{s}Y^*_{l'm} \mathfrak{h}_1 \, {}_{s}Y_{lm}.
\end{equation}
The functions ${}_{s}Y_{lm}$ are the spin-weighted spherical harmonics, and $d\Omega$ is the solid angle. It is easy to see in Eqs.~\eqref{spho_sphe} and~\eqref{clebsh} that one can recover the solution for the non-rotating (Schwarzchild) BH in the spherical harmonic basis when $a_f \rightarrow 0$.

\textit{Mode mixing} arises naturally from the choice of representation in PT in terms of spheroidal harmonics, as opposed to the spherical harmonics, which is the most natural representation in numerical relativity~(NR). 
Various authors~\cite{London_2018, Berti_2014} computed the values of spherical-spheroidal mixing coefficients, given by
\begin{equation}
    \sigma_{l'm',lmn}(a_f) = \delta_{m'm} \int_{\Omega} {}_{-2}Y^*_{l'm'}(\theta) {}_{-2}S_{lmn}(a_f\tilde{\omega}_{lmn};\theta) d\Omega.
\end{equation}
Since $\delta_{m'm}$ is the Kronecker delta parameter, one can drop the prime in the first $m$. Thus, one can find these coefficients in the literature, written without the first $m$ at all $\sigma_{l'lmn}$ or even written as $\mu_{ml'ln}$. With this representation, the strain takes the form 
\begin{equation}
    h_+(t) - ih_{\times}(t) = \sum_{l'} \sum_{lmn} h_{lmn}(t) \sigma_{l'mlmn}(a_f) {}_{-2}Y_{l'm}.
\end{equation}
In our case, the amplitudes and phases used in Eq.~\eqref{strain} belong to parametrizations made by London et al.~\cite{London_2020, London_2014}, where the mode mixing is already accounted for. Thus, we consider amplitudes labeled with $(l'mlmn)$ as
\begin{equation}\label{amp_sigma}
    A_{l'mlmn} = A_{lmn} \sigma_{l'mlmn}.
\end{equation}
Therefore, to consider the following three QNMs, namely $[(2,2,0), (3,3,0), (4,4,0)]$, we include $[(2,2,2,2,0), (3,2,2,2,0), (3,3,3,3,0), (3,3,3,3,0),\\
(4,4,4,4,0)]$, see Eq.~\eqref{amp_sigma}. Indeed, the resulting signal is a sum of decaying waves with amplitudes and phases for $lmn = [(2,2,0), (3,3,0), (4,4,0)]$. 


In consistency with current constrains on the fractional deviations \cite{ligo_testGR3} and the possible error on those parameters predicted for LISA \cite{toubiana2024} for a source with an SNR $\leq 600$, we inject fractional deviations of QNMs equal to $\delta \omega_{lmn}= [0.0, 0.01, 0.03]$ and $\delta\tau_{lmn} = [0.0, 0.05, 0.1]$ in the following order $lmn = [(2,2,0), (3,3,0), (4,4,0)]$. Of course, more QNMs could and should be added, but as a proof of concept, we decided to include only these three, leaving more complex searches for future work. 

We consider input data including a GW signal with and without noise. The sampling rate is set to 1 second as a compromise between the planned sampling rate of 0.25 s, the typical duration of the ringdown for a heavy source (about 7000 s for a mass of $10^7$ $M_\odot$) and the number of data points $8192$.
The parameters used for the source injection are listed in Table~\ref{tab:params_inj}. Note that we use $\iota$ as the inclination angle instead of $\theta$. We also write the ringdown SNR as well as the final parameters for the remnant BH, obtained with Eqs.~(3.6) and (3.8) from~\cite{PhenomD}.

\begin{ruledtabular}
\begin{table}
    \caption{Parameters for MBHB injection}
    \label{tab:params_inj}
    \centering
    \begin{tabular}{cc|cc}
        Parameter & Value & Parameter & Value \\
        \hline
         $m_1 (M_\odot)$& 9384087 & $m_2 (M_\odot)$ & 3259880  \\
         $\chi_1$ & 0.555 & $\chi_2$ & -0.525 \\
         $\iota$ (rad) &  $\pi/3$ & $\phi$ (rad) &  $\pi/4$ \\
         $\beta$ (rad) & $\pi/2$ & $\lambda$ (rad) & $\pi/3$ \\
         $D_l $(Mpc) & 50000 & q & 2.878\\
         $M_f (M_\odot)$ & 1.2175649 $\times 10^7$ & $a_f$ & 0.821\\
         $\delta \omega_{220}$ & 0.0 & $\delta \tau_{220}$ & 0.0\\
         $\delta \omega_{330}$ & 0.01 & $\delta \tau_{330}$ & 0.05\\
         $\delta \omega_{440}$ & 0.03 & $\delta \tau_{440}$ & 0.1\\
         SNR & 587\\
    \end{tabular}
\end{table}
\end{ruledtabular}

\subsection{Templates}\label{sec_template}

We consider two approaches where the recovery template takes different forms: the \textit{agnostic} approach and the \textit{deviations} approach.
In the \textit{agnostic} approach, we assume that the ringdown waveform is described by 
\begin{equation}
    h_+ - ih_{\times} = \sum_{k} A_{k} e^{i(\phi_{k} - t \Tilde{\omega}_{k})}.
\end{equation}
Here, complex frequencies, amplitudes, and phases can take any value. Note also that any dependence on the spheroidal harmonic is absorbed in the amplitude and the phase. In this way, no \textit{mode mixing} is specified. For example, in this approach, it would not be possible to know how much of the $(2,2,0)$ QNM contribution comes from the spherical harmonic $(2,2)$ or $(3,2)$. It differs from Eq.~\eqref{spheroidal} as no assumption is made on the value of the complex frequency nor the spherical contribution to any QNM. Despite the fixed number of modes~$k$, we can call this approach ``agnostic''.

In the \textit{deviations} approach, we assume the framework of GR as baseline but allow for a small ``deviation'' in the complex frequencies
\begin{equation}
    h_+ - ih_{\times} =  \sum_{lmn} A_{lmn} e^{-t/\tau'_{lmn} + i(\phi_{lmn} - t \omega'_{lmn}) },
\end{equation}
where $\omega'_{lmn}$ and $\tau'_{lmn}$ are the deviated frequency and damping time from Eqs.~\eqref{GR+dev}. In this case, one recovers GR when the deviations are zero. We impose which QNMs are present and look for each pair of deviations. We then compare the results from both approaches and discuss the information one can extract from them.

In our toy model, the injection and the recovery template have the same starting time. In this way, no error is introduced in the waveform due to the uncertainty of the ringdown starting time. Consequently, we do not try to evaluate any systematic uncertainties coming from the definition of the starting time of the ringdown, which is itself ill-defined~\cite{Bhagwat_2018, Bhagwat_2020, baibhav2023agnostic}. 
However, when dealing with real data, where one does not know the appropriate starting point, several starting times in the vicinity of the luminosity peak should be considered (see for example~\cite{Isi_2019, Giesler_2019, Carullo_2023}). We also fix the sky localization to the true value. Thus, no error from this parameter is introduced in the waveform either. We leave both issues to be explored in the future.

\subsection{\label{sec_bayes}Bayesian analysis}

The posterior distribution of the signal parameters $\boldsymbol{\theta}$ given the observed data $d$, in a Bayesian approach, is expressed as
\begin{equation}\label{posterior}
    p(\boldsymbol{\theta} \vert d, M) = \frac{p(d\vert \boldsymbol{\theta}, M) \,  p(\boldsymbol{\theta} \vert M)}{p(d \vert M)},
\end{equation}
where $p()$ are the probability density functions, $\boldsymbol{\theta}$ is the vector of the source physical parameters, $M$ is the model or any other feature considered. In the numerator, we have the likelihood $\mathcal{L}(\theta) =  p(d\vert \boldsymbol{\theta}, M)$ and the prior of the parameters  $\pi(\theta) = p(\boldsymbol{\theta} \vert M)$. 
In the numerator, $\mathcal{Z} = p(d \vert M)$ is the evidence, which is computed as the integral of the likelihood over the whole parameter's hyper-volume. 
For a noise with a covariance matrix $\mathbf{C}$, the likelihood takes the form
\begin{equation}\label{eq:log}
    \mathcal{L} =  \frac{1}{\sqrt{\text{det}(2\pi \mathbf{C})}}e^{-\frac{1}{2}(d -h(\boldsymbol{\theta}))\mathbf{C}^{-1}(d -h(\boldsymbol{\theta}))},
\end{equation}
whose logarithm can be written as
\begin{equation}
    \ln \mathcal{L} = -\frac{1}{2}(d - h(\boldsymbol{\theta})\vert d -h(\boldsymbol{\theta})) + \text{const} ,
\end{equation}
for a stationary noise. We drop the dependence on the model $M$ and we use the definition of the inner product in the time domain 
\begin{equation}\label{eq:inner_prod}
    (a \vert b) = \sum_{i,j=0}^{N-1} a_i(\boldsymbol{\theta})\mathbf{C}^{-1}_{ij} b_j(\boldsymbol{\theta}),
\end{equation}
where $N$ determines the time step $N*\Delta t = T$ of the total time.
We can then decompose the log-likelihood as 
\begin{equation}
\label{eq_lkh}
    \ln \mathcal{L} = (d \vert h(\boldsymbol{\theta}))  - \frac{1}{2} (h(\boldsymbol{\theta}) \vert h(\boldsymbol{\theta})) -\frac{1}{2} (d \vert d).
\end{equation}
The full log-likelihood is a sum over the log-likelihoods of the uncorrelated instrumental channels $A$ and $E$ (as we ignore the channel $T$ in this analysis). Then, we can write 
\begin{equation}
\label{eq_lkh_tdi}
    \ln \mathcal{L} = \sum_{I=A,E} \ln \mathcal{L}_{I}.
\end{equation}

\begin{ruledtabular}
\begin{table}
    \caption{Priors on the parameters estimated for both approaches}
    \label{tab:priors}
    \centering
    \begin{tabular}{ccc}
        Parameter &  Priors for \textit{agnostic} & Priors  for \textit{deviations} \\
        \hline
        $A_{lmn}$ & log uniform [-23,-16] & log uniform [-23,-16]\\
        $\phi_{lmn}$ & uniform [0,$2\pi$] & uniform [0,$2\pi$]\\
        $\omega'_{lmn}$ & uniform [$10^{-5}$,0.1] & - \\
        $\tau'_{lmn}$ & uniform [1,$10^{5}$] & - \\
        $\delta\omega_{lmn}$ & - & [-0.2, 0.2] \\
        $\delta\tau_{lmn}$ & - & [-0.2, 0.2] \\
        $M_f$ & - & [0.9, 1.1] $\times M_f$\\
        $a_f$ & - & [0.9, 1.1] $\times a_f$
    \end{tabular}
\end{table}
\end{ruledtabular}

To obtain the covariance matrix, we use the same method as in~\cite{Isi_21} with an analytical power spectral density (PSD). Namely, one can create the covariance matrix as a symmetric Toeplitz matrix, assuming stationarity, such that 
\begin{equation}
    \mathbf{C}_{ij} = \rho(|i-j|),
\end{equation}
where $\rho (k)$,  $k=|i-j|$, is the autocovariance function (ACF) that can be estimated from noise-only data in the time domain with a length longer than $N$ or as the inverse Fourier transform of the PSD. In our case, we use the latter option, generating the ACF from the LISA Science Requirements Document (SCiRD) \cite{SciRD} PSD 
\begin{equation}
    \rho (k) = \frac{1}{2T} \sum_{j=0}^{N-1} S(|f_j|)e^{2\pi ijk/N}.
\end{equation}

\begin{figure*}
    \centering
    \includegraphics[width=0.9\textwidth]{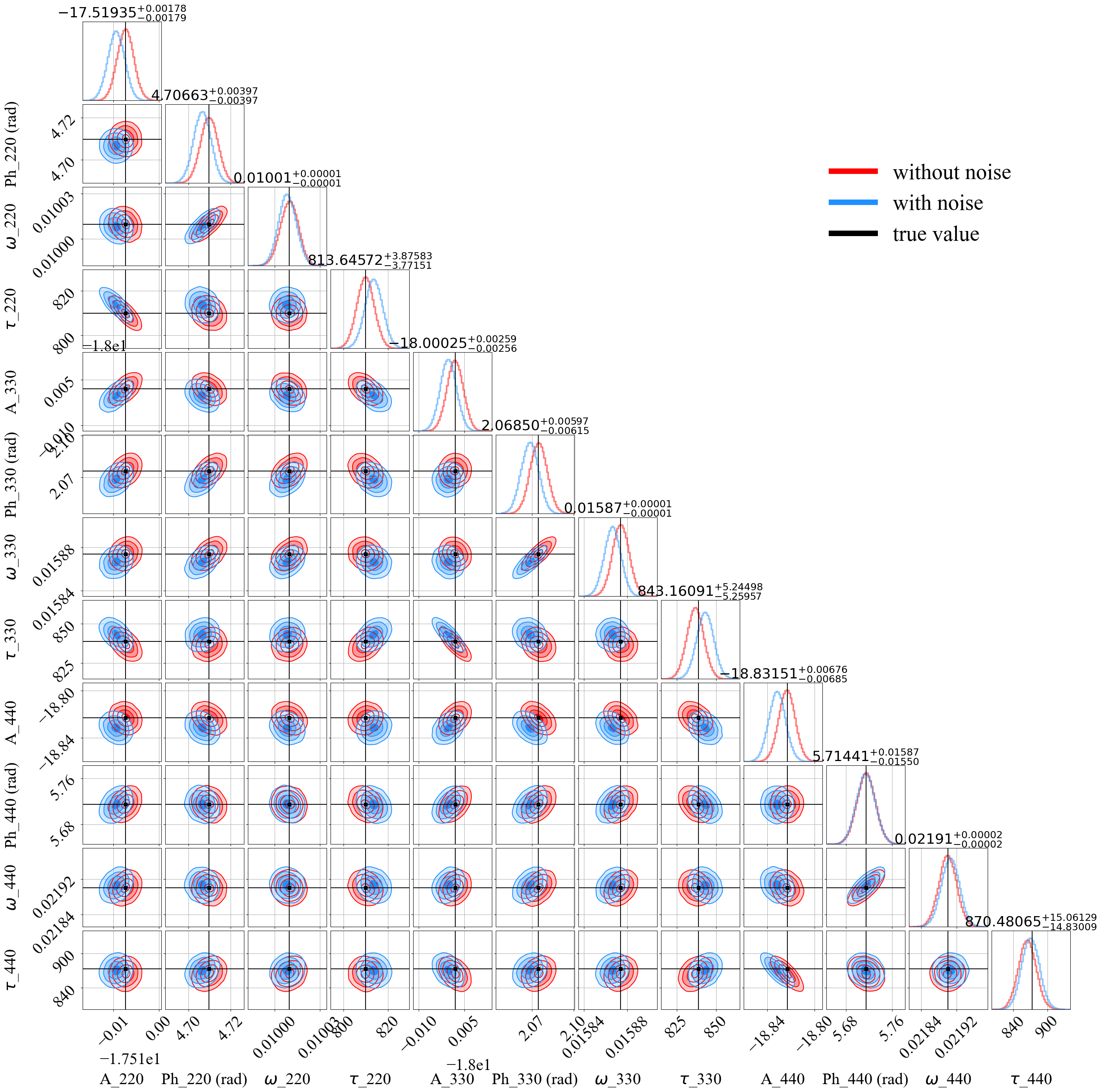}
    \caption{Posterior distribution for the \textit{agnostic} case with 4 dimensions per mode \{$A_{k}, \phi_{k}, \omega_k, \tau_k$~\}. Posterior distribution without noise injection in red, with noise injection in blue and injected values marked with black lines. Overall the distributions agree with the true values, with some fluctuations in the noisy case as expected.} 
    \label{fig:ag_res11}
\end{figure*} 

Working with matrices usually requires long computational time and special care must be taken because of their numerical instability. To reduce the computational time, one can use different methods such as the Cholesky decomposition \cite{Cholesky} or the Levinson recursion \cite{levinson, Durbin} among others. To compute fast inner products, we use the \texttt{bayesdawn} package~\cite{Bayesdawn}. More accurately, we make use of the implemented preconditioned conjugate gradient (PCG)~\cite{KAASSCHIETER_1988} and the Jain method~\cite{Jain} to avoid numerical errors and to fast compute the values of the vectors 
\begin{equation}
    \overline{a}_j (\boldsymbol{\theta}) = a_i(\boldsymbol{\theta})\mathbf{C}^{-1}_{ij}.
\end{equation}
Then, the inner product of Eq.~\eqref{eq:inner_prod} becomes a fast product of vectors 
\begin{equation}
    (a \vert b) = \sum_{j=0}^{N-1} \overline{a}_j (\boldsymbol{\theta}) b_j(\boldsymbol{\theta}).
\end{equation}

\begin{figure}[t]
    \centering
    \includegraphics[width=0.45\textwidth]{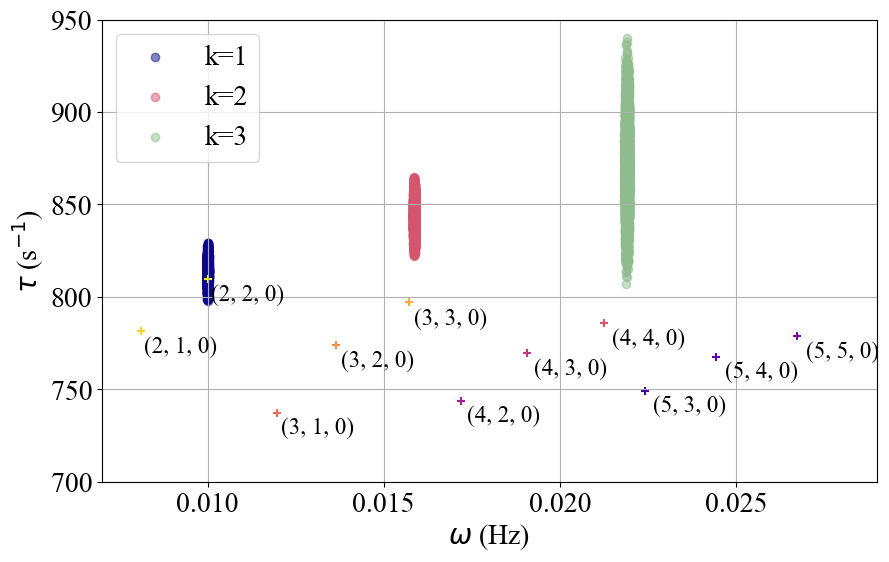}
    \caption{Posterior distribution of the pairs of complex frequencies in the spectrum map. Each mode `k' is associated with one color, purple, pink, or green. The spectrum of the true BH is represented with colored crosses with their QNM label nearby.}
    \label{fig:ag_wtau}
\end{figure}

\section{Results}\label{sec4}

As mentioned in Sec.~\ref{sec_template}, we consider two analysis approaches. For each approach, we perform two runs: with and without noise. In the following, we discuss the results with noisy data obtained with the \texttt{dynesty}~\cite{dynesty} sampler. 

\subsection{Agnostic approach} 

In the \textit{agnostic} approach the parameters are $\boldsymbol{\theta}~=~\{ A_k, \phi_k, \omega_k, \tau_k\}$ with $k=1,2,3$ accounting for the three QNMs $[(2,2,0), (3,3,0), (4,4,0)]$. To avoid any degeneracy between the modes, we impose the condition of hyper-triangulation in the frequency. This condition restricts the second frequency to be larger than the first one, and the third to be larger than the second. Then, the uniform prior of each frequency decreases relative to the previous one, like an inverted triangle in the prior volume. The amplitudes have a logarithmic uniform prior in $[-23,\, -16]$, while the frequency has a uniform prior in $[10^{-5},\, 0.1]$ and the damping time in $[1,\, 10^{5}]$. The phase is allowed to take any value in the range~$[0,\, 2\pi]$. These values can also be found in Table~\ref{tab:priors}.

\begin{figure}
    \centering
    \includegraphics[width=0.4\textwidth]{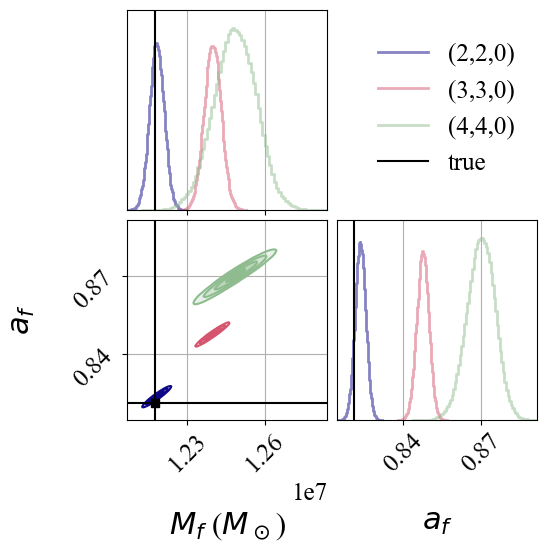}
    \caption{Posterior distribution for the mass and spin computed with Eq.\eqref{fittings} for each pair of ($\omega_{lmn}, \tau_{lmn}$). Note the agreement of $(2,2,0)$ with the true value while the other two modes diverge, showing a deviation from GR in those modes.}
    \label{fig:ag_maf}
\end{figure}

In Fig.~\ref{fig:ag_res11}, we present the posterior distribution of the injection without noise in red, with noise in blue, and the injected values with black lines. Remember that the values of the deviations injected in the waveform have been introduced in Table~\ref{tab:params_inj}. We show 12 parameters, 4 for each QNM. In general, the Gaussian distributions converge to the true values, with some minor fluctuations in the noisy case, as expected. In the following, we will discuss the results from the noisy dataset only. In the figure we have already labeled the name of the QNMs ``$k$" since we know them from the injection. However, we will not know which modes are present in the future LISA data analysis. Therefore, one must first find the QNM corresponding to each label $k=lmn$. 

One could identify the QNMs by comparing the values of the complex frequencies ($\omega_k$, $\tau_k$) with pairs of ($\omega_{lmn}$, $\tau_{lmn}$) corresponding to an assumed mass and spin obtained from an IMR analysis carried out beforehand. We present the idea of this approach in Fig.~\ref{fig:ag_wtau}. The scatter points correspond to the values of the posterior distribution for $k=1$ in purple, $k=2$ in pink, and $k=3$ in green. The colored crosses correspond to the values of ($\omega, \tau$) easily identified with the QNMs labels written nearby. By looking at this figure, we can already state that there might be a deviation from GR, as there is only one mode that we can confidently identify with the posterior distributions, that is $k:1 = (2,2,0)$. The other two clusters of points could be assigned to their nearest QNM, namely  $k:2 = (3,3,0)$ and $k:3 = (4,4,0)$. At this stage, one could make one of the two following hypotheses:
\begin{enumerate}[label=(\roman*)]
    \item The IMR estimation is trustworthy and the final mass and spin are taken to be the true values. In this case, the dominant mode could exhibit deviations from GR as well as all the other harmonics.
    \item The IMR estimation on mass and spin itself can have systematic errors. Therefore, the analysis should be done by relying only on QNMs. We can identify a QNM that does not present a deviation of GR and assume the inferred mass and spin from that QNM as the true value.
\end{enumerate}

\begin{figure*}
    \centering
    \includegraphics[width=1\textwidth]{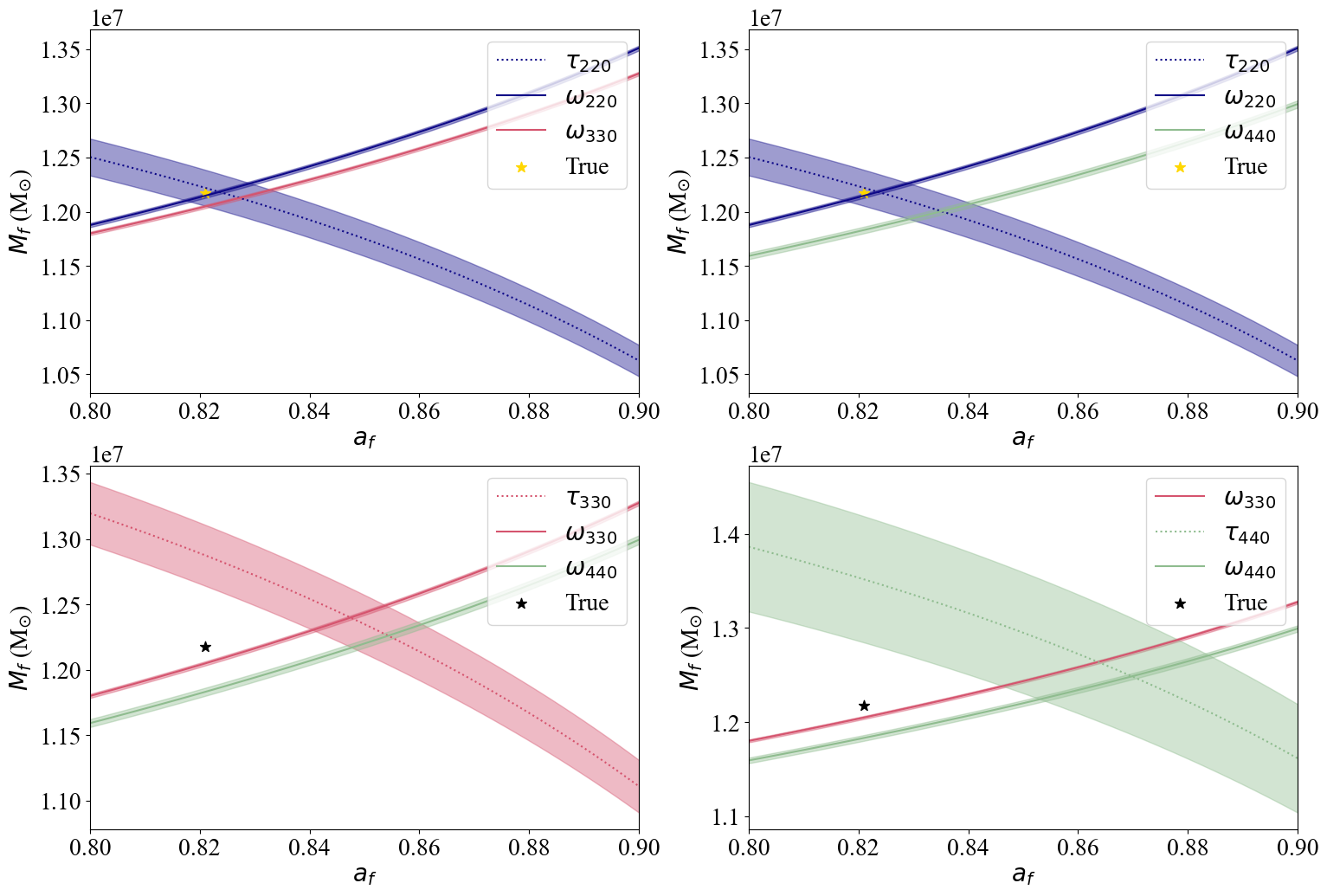}
    \caption{Computed mass for spins in the range [0.8-0.9] with Eq.~\eqref{fittings} with a $99\%$ confidence level, from the estimated mean values of ($\omega_{lmn}, \tau_{lmn}$)}
    \label{fig:ag_gossan}
\end{figure*}

One possible way to check for consistency between QNMs is to trace back the mass and spin of the remnant BH, using for example the parametrization from~\cite{Berti_2006}:
\begin{subequations}\label{fittings}
\begin{align}
    M \omega_{lmn}  =  & f_1 + f_2 (1 - j)^{f_3},\\
    \omega_{lmn}\tau_{lmn} / 2 = & q_1 + q_2(1-j)^{q_3},
\end{align}
\end{subequations}
with $f_1, f_2, f_3, q_1, q_2, q_3$ fitting parameters from Tables (VIII, IX, X) in~\cite{Berti_2006}. With any pair of~($\omega, \tau$) one can compute first the value of the spin and then the value of the mass. 

If we take samples within each mode's posterior distribution in Fig.~\ref{fig:ag_wtau} and use Eqs.~\eqref{fittings} to compute the corresponding masses and the spins, we obtain the distributions in Fig.~\ref{fig:ag_maf}. We perceived already from Fig.~\ref{fig:ag_wtau} that the posteriors of the mass and spin obtained from different QNMs would not overlap completely. Here we can confirm it by observing three different mean values for the spin without any overlap and three distributions for the mass with overlap between values computed from $(3,3,0)$ in pink and $(4,4,0)$ in green. Notably, the true value in black does not fit perfectly with the mean value of the $(2,2,0)$ in purple. This is due to small fluctuations in the ($\omega,\tau$) mean value, which can be seen in Fig~\ref{fig:ag_res11}, and the fact that Eqs.~\eqref{fittings} come from a fitting and thus, intrinsic errors of the order of $\sim 1-3 \%$ \cite{Berti_2006} are expected in the mass and the spin.

\begin{figure*}[t]
\begin{subfigure}[b]{0.3\textwidth}
    \centering
    \includegraphics[width=1\textwidth]{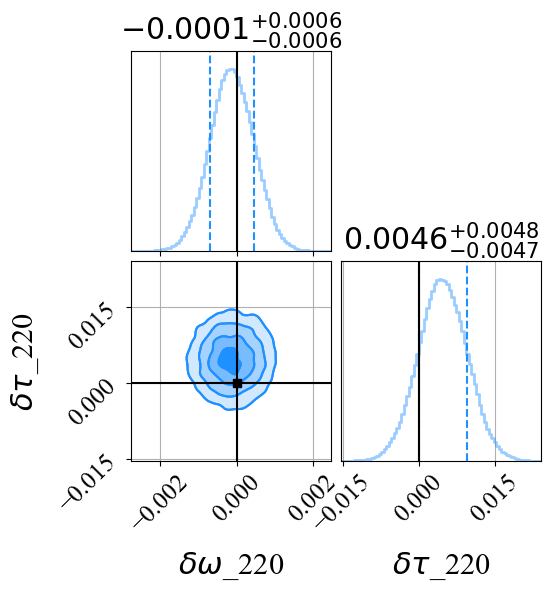}
    \caption{Fractional deviation obtained in $\omega_{220}$ and $\tau_{220}$}
    \label{fig:ag2_220}
\end{subfigure}
\begin{subfigure}[b]{0.3\textwidth}
    \centering
    \includegraphics[width=1\textwidth]{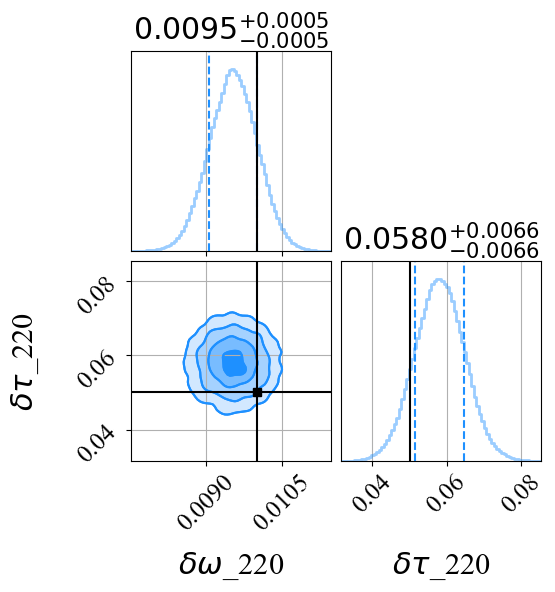}
    \caption{Fractional deviation obtained in $\omega_{330}$ and $\tau_{330}$}
    \label{fig:ag2_330}
\end{subfigure}
\begin{subfigure}[b]{0.3\textwidth}
    \centering
    \includegraphics[width=1\textwidth]{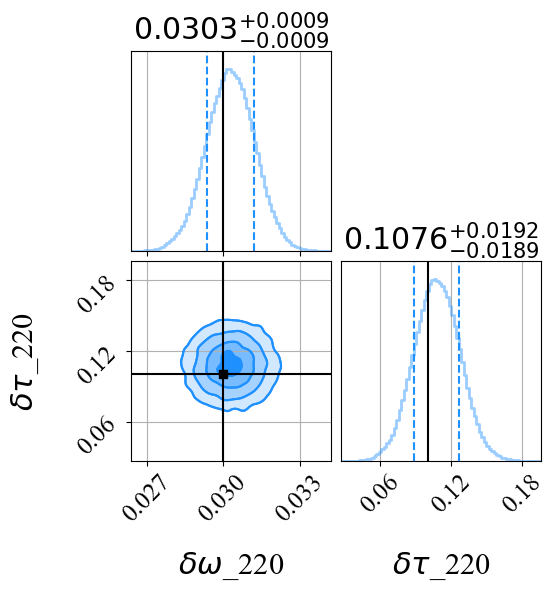}
    \caption{Fractional deviation obtained in $\omega_{440}$ and $\tau_{440}$}
    \label{fig:ag2_440}
\end{subfigure}
\caption{Posterior distribution of the fractional deviations in the complex frequency obtained from the posterior distribution of $\omega_{k}, \tau_{k}$, compared to the GR QNMs with true values of $M_f, a_f$, for the modes $k=lmn$. }
\label{fig:ag2}
\end{figure*}

To better understand the differences in the posteriors, one could alternatively follow the approach adopted in ~\cite{Gossan_2012}. That is, using Eqs.~\eqref{fittings} to compute the value of the mass for a given spin and comparing the values obtained from different QNMs. Note, that this approach does not propagate errors from the spin fitting into the mass, as it remains fixed. This representation can be seen in Fig~\ref{fig:ag_gossan}, where the true value is marked with a golden or a black star, and the shadow lines correspond to the $99\%$ credible levels. The standard deviation for the mass is related to the standard deviations of $\omega$ and $\tau$. As a result of the precision on the frequencies posteriors, the uncertainty bands derived from $\omega_{lmn}$ are relatively narrow. The mass and spin obtained from the $(2,2,0)$ mode are consistent with the injected value, while the others exhibit deviations from it.

\subsubsection{Using the IMR masses and spins as references}

We can now discuss the first hypothesis (\textit{i}) stated above. Imagine we want to quantify the deviation in each mode's frequency to put some constraints on an alternative theory. In that case, we have to compare the posterior distributions of frequency and damping time with the QNM values for a BH with respect to the IMR estimated final mass and spin. To simplify, we assume that the parameters estimated from the IMR analysis equal the exact injected values. Results can be seen in Fig.~\ref{fig:ag2}, where we subtract the true complex frequency value from each mode. In this figure, we can see that each posterior agrees with the injected value within $2 \sigma$. The dashed blue lines mark the quantiles $(0.16,\, 0.84)$, i.e., the $1 \sigma$ distribution. 

However, using reference values for the mass and spin is somewhat inconsistent with the agnostic philosophy. Keep in mind that, the mean values estimated from an IMR analysis could present a bias. Moreover, using the whole parameter posterior distribution instead of mean values would better allow for the propagation of uncertainties. 

\begin{figure}
    \centering
    \includegraphics[width=0.45\textwidth]{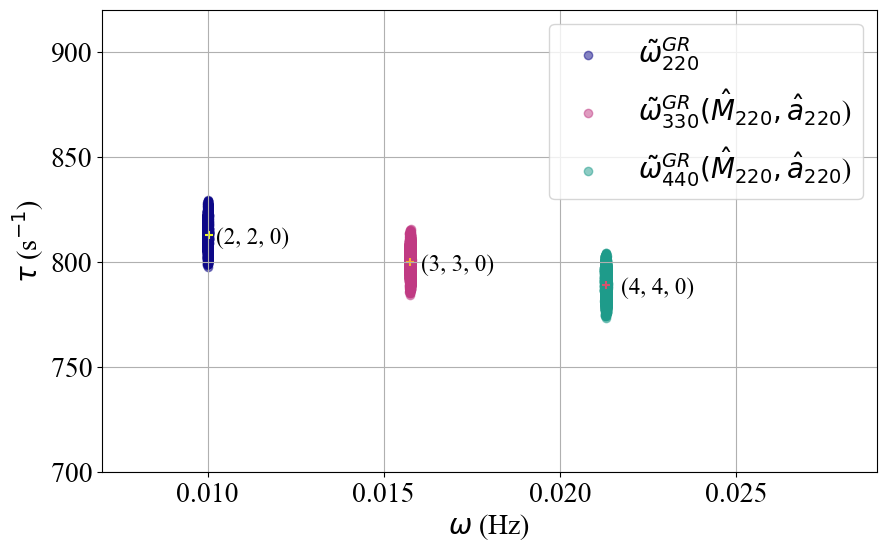}
    \caption{Posterior distribution of each mode, generated from the posterior distribution of mass and spin derived from the (2,2,0) mode, compared with true spectrum.}
    \label{fig:ag_wtau_maf}
\end{figure}

\subsubsection{Relying only on QNM characterization}

Without a posterior distribution from an IMR waveform inference, we now adopt the second of the two above hypotheses and use the mass and spin obtained from the dominant mode as reference values. While we already showed above that the dominant mode agrees within a $99\%$ credible confidence with the injected parameters, there is a risk in assuming that the $(2,2,0)$ mode does not deviate from GR. Correspondingly, deviations in the dominant mode might also appear. 

Now, to quantify the deviations in this framework, we should translate the differences in mass and spin into deviations in $\omega$ and $\tau$ in terms of $\delta \omega, \delta \tau$, see Eqs.~\eqref{GR+dev},~\eqref{dev}. To this end, we assumed that the posterior distribution obtained from the $(2,2,0)$ mode is the ``true" description of the remnant BH in GR. We can then compute the QNM spectrum with the derived mass and spin from that mode. This computation is shown in Fig~\ref{fig:ag_wtau_maf}, where we observe the posterior of the three complex frequencies $\tilde{\omega}_{lmn}$ computed with the mass and spin derived from the dominant mode $\{\hat{M}_{f220}, \hat{a}_{f220}\}$. The GR values are marked with colored crosses on top of the distributions. 

\begin{figure*}[t]
\begin{subfigure}[b]{0.49\textwidth}
    \flushleft
    \includegraphics[width=1.04\textwidth]{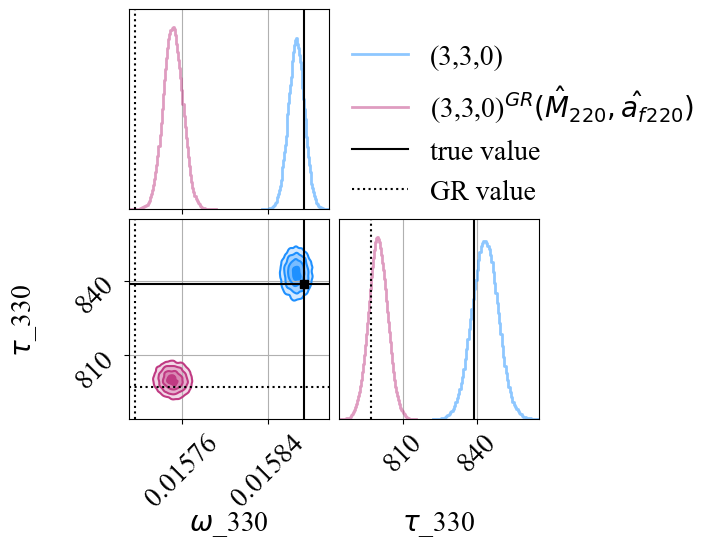}
    \caption{Posterior distribution of $\omega_{330}$ and $\tau_{330}$ against the distribution in the GR framework.}
    \label{fig:ag_wt330}
\end{subfigure}
\begin{subfigure}[b]{0.49\textwidth}
    \flushright
    \includegraphics[width=1.04\textwidth]{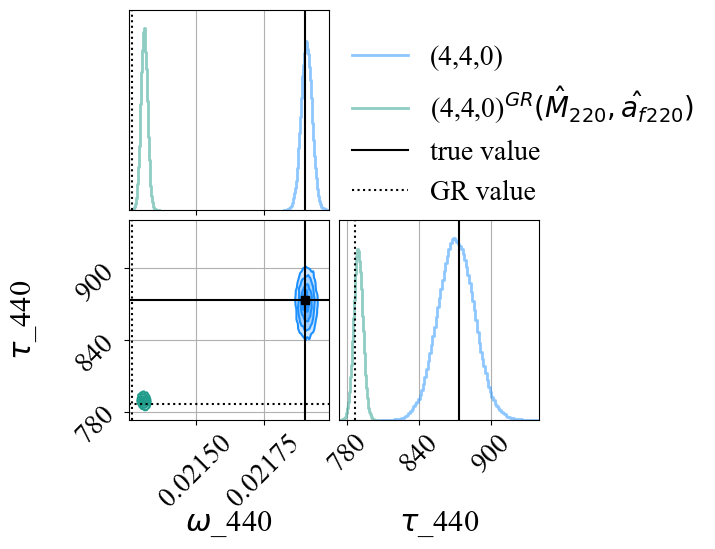}
    \caption{Posterior distribution of $\omega_{440}$ and $\tau_{440}$ against the distribution in the GR framework.}
    \label{fig:ag_wtau440}
\end{subfigure}
\begin{subfigure}[b]{0.44\textwidth}
    \centering
    \includegraphics[width=1\textwidth]{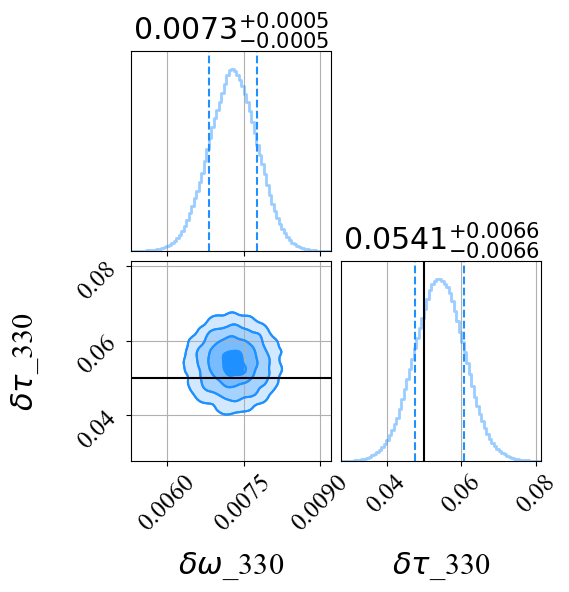}
    \caption{Fractional deviation from GR for $\tilde{\omega}_{330}$.}
    \label{fig:ag_d330}
\end{subfigure}
\begin{subfigure}[b]{0.44\textwidth}
    \centering
    \includegraphics[width=1\textwidth]{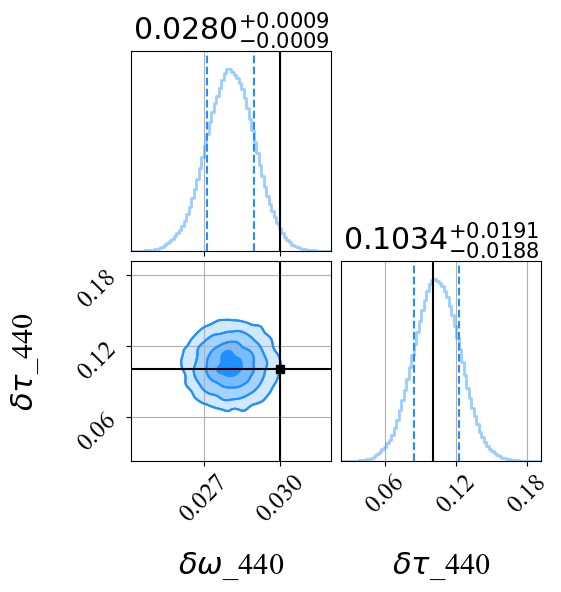}
    \caption{Fractional deviation from GR for $\tilde{\omega}_{440}$.}
    \label{fig:ag_d440}
\end{subfigure}
\caption{Evidence of deviations in the frequency and damping time computed with the estimated final mass and spin. The top row shows the posterior distribution of the complex frequency for each mode in blue, against the computed posterior distribution in the GR framework for the parameters derived from the (2,2,0) modes in pink and green. In the bottom row, we subtract the mean value of the estimated GR QNM from the obtained posterior distribution. By doing so, the fractional deviation becomes evident. The damping time agrees with the injected value (black lines) for both modes, while the frequency presents a bias due to the high sensitivity to the remnant parameters estimated from the (2,2,0) mode.}
\label{fig:ag_devs}
\end{figure*}

Given the distributions without deviations, one can measure the one-dimensional deviation for each parameter. To this aim, we need to compare the complex frequency $(\tilde{\omega}_{lmn})$ posteriors with the QNM values obtained from the mass and spin corresponding to the $(2,2,0)$ mode $(\hat{M}_{f220}, \hat{a}_{f220})$ for each mode. This is analogous as comparing Fig.~\ref{fig:ag_maf} with Fig.~\ref{fig:ag_wtau_maf}. The results are shown in the top row of Fig.~\ref{fig:ag_devs}, where we find the distribution of GR complex frequencies computed with the mass and spin derived from the $(2,2,0)$ mode for the $(3,3,0)$ mode in pink, and in green for the $(4,4,0)$ mode. We can easily distinguish them from the non-GR values obtained from the sampler in blue.

A simple equation to quantify this tension is commonly used \cite{leizerovich2023tensions, Lemos_2021}:
\begin{equation}\label{eq:sigma_dev}
    N_{\sigma} = \frac{|\mu_A - \mu_B |}{\sqrt{\sigma_A^2 + \sigma_B^2}},
\end{equation}
where A and B are two different models, $\mu$ is the estimated mean value and $\sigma$ is the standard deviation. This equation gives the number of standard deviations between two posterior distributions in one dimension. This simple definition can be used as a means to estimate uncertainties in the following. For the injected values of Table~\ref{tab:params_inj}, the computed standard deviation from GR values is shown in Table~\ref{tab:st_dev}. Should this be observed, we would have detected a deviation from GR in $\omega_{330}$ with more than 10 standard deviations relative to the $(2,2,0)$ mode. It is also important to note that even though we distinguished a deviation from GR with high precision, the injected GR value does not correspond to the recovered GR value, thus revealing a bias. 

\begin{ruledtabular}
\begin{table}[]
\caption{Computed uncertainty from GR for the injected parameters in the agnostic case.}
    \centering
    \begin{tabular}{c c c}
    &$N_{\sigma^{GR}}$(3,3,0) & $N_{\sigma^{GR}}$(4,4,0) \\
    \hline
     $\delta \omega$   & 10.31 & 28.46  \\
     $\delta \tau$    & 7.97 & 5.62
    \end{tabular}
    \label{tab:st_dev}
\end{table}
\end{ruledtabular}

To compute the values of the fractional deviations for each mode, one can use the blue posterior distributions from the top row and subtract the mean value of the pink and green posteriors respectively. Performing such computation leads to the figures in the bottom row in Fig.~\ref{fig:ag_devs}. As expected from the previous remark on the GR inconsistent values, the injected parameter does not perfectly agree with the posterior distribution. Indeed, the estimated value shown in Fig.~\ref{fig:ag_d330}, is inconsistent with $\delta\omega_{330} = 0.01$. This is because we used the GR value derived from the mass and spin inferred from the $(2,2,0)$ mode characterization, for which we assumed no deviation. Even if the mass and spin computed from the $(2,2,0)$ mode agree with the true values, the assumption of no deviation in this mode might have strong implications, as any fluctuation on the $(2,2,0)$ mode will translate into fluctuations in the estimated mass and spin and therefore in the subsequent characterization of the $(3,3,0)$ and $(4,4,0)$ modes. Certainly, the computation of the QNMs highly depends on the mass and spin, thus small variations of those intrinsic parameters translate to larger variations on the complex frequency parameter space. To summarise, with the hypothesis (\textit{ii}), this kind of analysis would allow one to differentiate GR from another theory. However, when attempting to constrain an alternative theory, the recovered values of the deviations might lead to misinterpretations.

One can avoid this type of discrepancy by using the posterior distribution of the mass and spin inferred from the full IMR instead of the posterior inferred from the $(2,2,0)$ mode. Again, this implies that the IMR analysis should provide unbiased values. Keep in mind that, in the present analysis, the mass and spin from the $(2,2,0)$ mode were consistent within $2 \sigma$ with the injected values of final mass and final spin. Nevertheless, deviations might appear in the $(2,2,0)$ mode. Thus, comparing the mass and spin estimated from this BH spectroscopy with those inferred from the full IMR waveforms would be informative. 

\begin{figure*}
    \centering
    \includegraphics[width=0.9\textwidth]{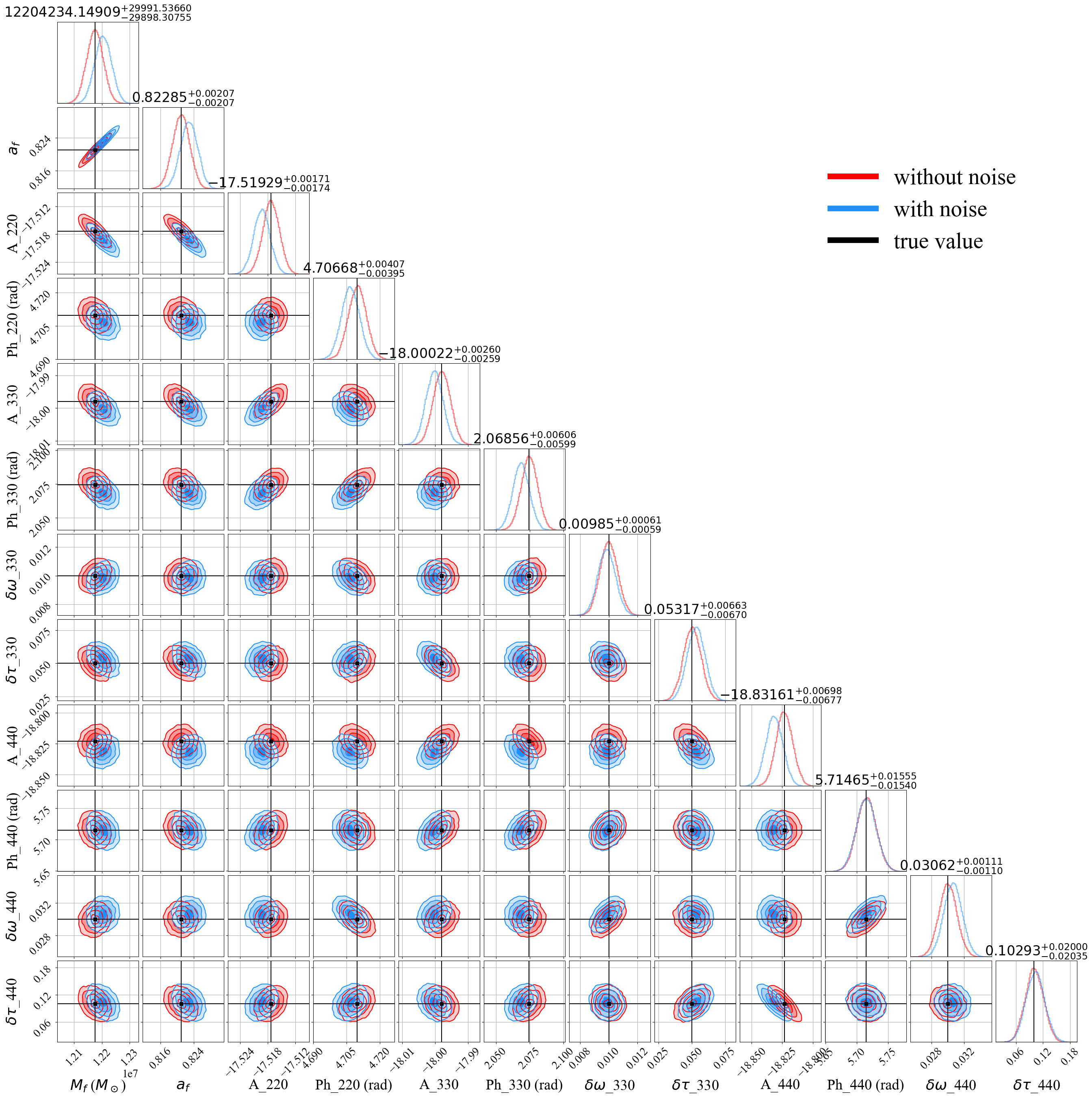}
    \caption{Posterior distribution for the \textit{deviations} case. Results without noise injection in red, with noise injection in blue, and injected value with black lines. Overall the distributions agree with the true values, with some fluctuations in the noisy case as expected.}
    \label{fig:dev_res04}
\end{figure*}

\subsection{Deviations approach}

\begin{figure*}[t]
\begin{subfigure}[b]{0.45\textwidth}
    \centering
    \includegraphics[width=1\textwidth]{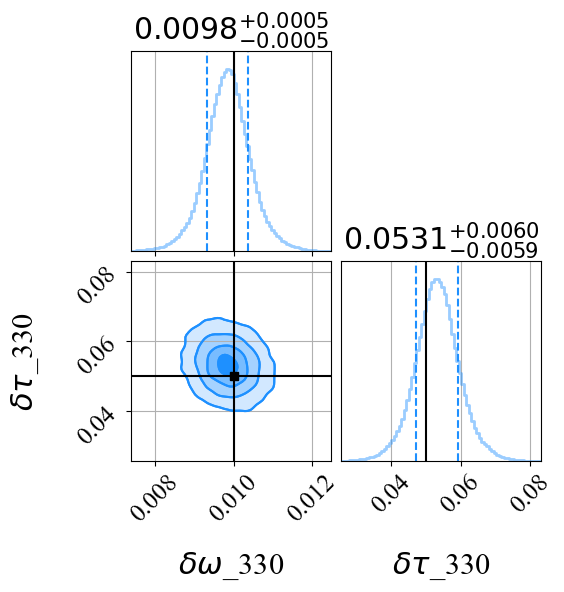}
    \caption{Fractional deviations in (3,3,0)}
    \label{fig:dev_330}
\end{subfigure}
\begin{subfigure}[b]{0.45\textwidth}
    \centering
    \includegraphics[width=1\textwidth]{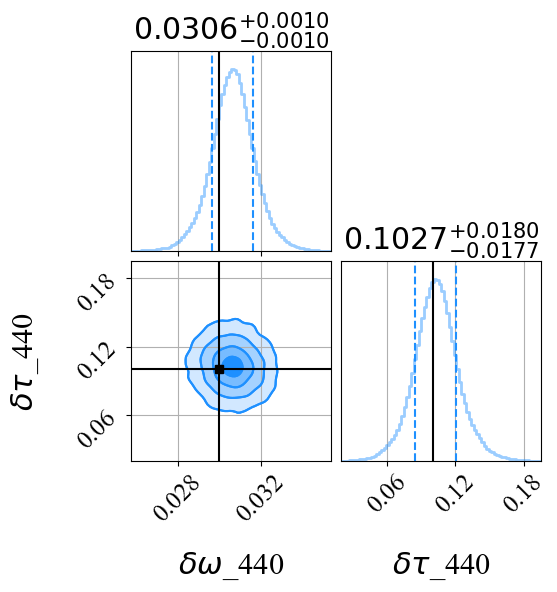}
    \caption{Fractional deviations in (4,4,0)}
    \label{fig:dev_440}
\end{subfigure}
\caption{Posterior distribution of fractional deviations in modes (3,3,0) and (4,4,0) directly from the sampler. Dashed lines denote the 1$\sigma$ error and black lines the injected value. We obtain agreeing results with high precision.}
\label{fig:devs}
\end{figure*}

In the following, we discuss the results of the second approach, the \textit{deviations} template. For this search, we have to define beforehand which QNMs appear in the waveform. We also assume that the dominant mode does not have deviations from GR. Imposing this condition allows us to break the degeneracy between the mode's fractional deviations from GR and the BH mass and spin. Alternatively, one could fix the mass and the spin, and allow the whole QNM spectrum to present deviations.

The parameters in this second approach are $\boldsymbol{\theta}~=~\{ M_f, a_f, A_{220}, \phi_{220}, A_{k}, \phi_{k}, \delta \omega_k, \delta \tau_k\}$ with k=$[(3,3,0), (4,4,0)]$. The mass and spin have a uniform prior within a range of $10 \% $ around the injected value, which gives the intervals $[0.9,\, 1.1] \times M_f$ and $[0.9,\, 1.1] \times a_f$, respectively. The values of the priors can be found in Table~\ref{tab:priors}. The phase has a uniform prior in the range $[0,\, 2\pi]$, while the amplitudes have a logarithmic uniform prior in $[-23,\, -16]$. Lastly, the deviations have a uniform prior in the range $\delta\omega, \delta\tau = [-0.2,\, 0.2]$. This range arises naturally from the chosen QNMs, as the relative difference between two QNMs is bigger than $0.2$:
\begin{equation}
    \frac{|\omega_{220} - \omega_{330}|}{\omega_{220}} >0.2.
\end{equation}
For QNMs with closer spectrum such as $(2,2,0)$ and $(2,2,1)$, there is a possible switching on the labels, producing a degeneracy between those two modes and exhibiting a bimodal posterior distribution. For this reason, we do not include the QNM $(2,2,1)$ in the analysis, even though its presence might have been detected in GW150914~\cite{Isi_2019} albeit the small significance (see the discussion in~\cite{Isi_2019, Giesler_2019, cotesta_22, isi2022revisiting, Isi_2023, Carullo_2023}). We leave this particular case to be studied in the future.

In Fig.~\ref{fig:dev_res04} we show the posterior distribution with and without noise injection in blue and red respectively. Injected values are marked with black lines. We observe the consistency between both results with the true values. In the following, we analyze the results from the dataset with noise injection.

Note that in this approach, the analysis is straightforward. The fractional deviations in the spectrum directly result from the posteriors since the deviations found in each QNM account for the estimated mass and spin by construction. In Fig.~\ref{fig:devs} we zoom in on the deviations of the $[(3,3,0), (4,4,0)]$ modes and recover the injected values with high accuracy and precision. The uncertainty on the deviations from GR parameters $\delta \omega$ and $\delta \tau$ with this template are listed in Table~\ref{tab:st_dev_dev}. Note that under the same hypothesis ($ii$) as in the previous analysis, i.e. no deviation in the ($2,2,0$) is observed, it is possible to derive constraints on an alternative theory, since the injected values are within the posterior distributions. A caution message is imperative here. The template considered, by construction, does not allow for deviations in the dominant mode. The effect of a fractional deviation in the ($2,2,0$) mode, when not considered in the search template needs further investigation. Nevertheless, to constrain an alternative theory the model-independent template might not be enough, and specific templates for beyond-GR theories are required.


\begin{ruledtabular}
\begin{table}[t]
\caption{Computed deviation uncertainty from GR for the injected parameters in the \textit{deviations} approach.}
    \centering
    \begin{tabular}{c c c}
    &$N_{\sigma^{\mathrm{GR}}}$(3,3,0) & $N_{\sigma^{\mathrm{GR}}}$(4,4,0) \\
    \hline
     $\delta \omega$   & 16.34 & 27.81  \\
     $\delta \tau$    & 7.98 & 5.06
    \end{tabular}
    \label{tab:st_dev_dev}
\end{table}
\end{ruledtabular}

Given that the value of the standard deviation for each parameter is inversely proportional to the SNR, there is a way to estimate the SNR needed to observe a specific deviation from GR with a given uncertainty in terms of standard deviation. We expand on this idea in Sec.~\ref{sec_GR_SNR}.


\begin{figure*}[t]
\begin{subfigure}[b]{0.45\textwidth}
    \centering
    \includegraphics[width=1\textwidth]{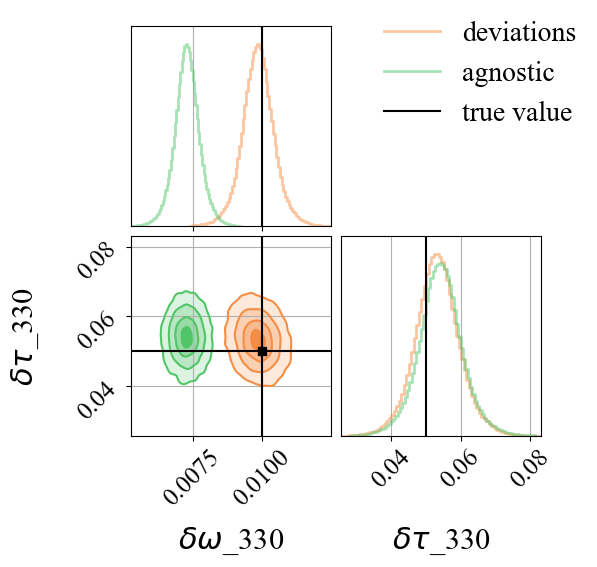}
    \caption{Fractional deviation for (3,3,0) mode with both methods}
    \label{fig:dev_ag330}
\end{subfigure}
\begin{subfigure}[b]{0.45\textwidth}
    \centering
    \includegraphics[width=1\textwidth]{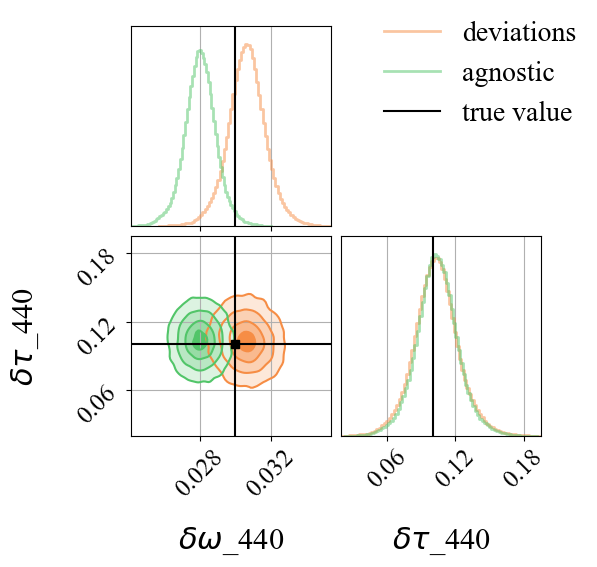}
    \caption{Fractional deviation for (4,4,0) mode with both methods}
    \label{fig:dev_ag440}
\end{subfigure}
\caption{Comparison of the posterior distribution of the fractional deviations for modes (3,3,0) and (4,4,0) for the two different methods. Posterior distributions obtained for the \textit{deviations} approach are shown in orange and green for the \textit{agnostic} approach. The black lines intersections mark the injected values.}
\label{fig:dev_ag}
\end{figure*}

\subsection{Discussion}\label{sec_disc}

In the perspective of testing the no-hair theorem and possible deviations from GR with the LISA instrument, we explore the extent to which we can extract the largest amount of information through two different analyses in terms of two generic templates. One possible approach is to compare the posterior distribution of fractional deviations in frequency and damping times $\delta \omega_k, \delta \tau_k$ from different templates. 

We compare both methods in Fig.~\ref{fig:dev_ag}, where the posterior distributions of deviations for the \textit{agnostic} approach are shown in green and the results for the \textit{deviations} approach are in orange. We denote the injected values by black lines. Under the same assumption of no deviation from GR in the ($2,2,0$) mode, the deviation approach gives more accurate results, making it possible to constrain alternative theories to GR. In the agnostic result, the premise that no deviation from GR affects the $(2,2,0)$ mode has strong implications. If one relaxes this constraint and assumes that the IMR estimation is accurate enough to fix the mass and the spin values, then a deviation in the dominant mode can be considered and the deviations of higher harmonics would be consistent with the injected values, as seen in Fig.~\ref{fig:ag2}. However, this result will strongly depend on the estimated mass and spin from the full IMR analysis, whose values can be biased if features like higher harmonics, eccentricity, or precession, to name only a few, are not considered.

\section{\label{sec_GR_SNR}Test of GR versus SNR}

In what follows, we discuss the uncertainty in the fractional deviations for a source with different SNR across the LISA band. To this aim, we will use the \textit{deviations} template, which provides the best consistency under the assumptions taken. We compute the standard deviation
\begin{equation}\label{eq:sigma}
    \sigma_{\theta_i} = \sqrt{\Gamma_{ii}^{-1}},
\end{equation}
where $\Gamma_{ii}$ is the Fisher matrix computed as the inner product defined in Eq.\eqref{eq:inner_prod} of the partial derivatives with respect to the parameter $\boldsymbol{\theta}_i$
\begin{equation}
    \Gamma_{ii} = \left(\frac{\partial a(\boldsymbol{\theta})}{\partial \boldsymbol{\theta}_i} \Big{|} \frac{\partial b(\boldsymbol{\theta})}{\partial \boldsymbol{\theta}_i}\right).
\end{equation}

The results are presented in Table~\ref{tab:FM}, showing the consistency between the error obtained from the Bayesian analysis with the Fisher forecast. Even if the Fisher matrix underestimates the uncertainty for $\delta\tau_{440}$, possibly due to the noise injection, we can still extract information from the other parameters.

\begin{ruledtabular}
\begin{table}[b]
    \caption{Uncertainty computed with the Fisher matrix (second column) and obtained with the sampler (third column).}
    \centering
    \begin{tabular}{ccc}
    & $\sigma_{\mathrm{FM}}$   & $\sigma_{\mathrm{sampler}}$  \\
    \hline
    $\delta \omega_{330}$  & $0.587 \times 10^{-3}$ & $0.602 \times 10^{-3}$\\
    $\delta \tau_{330}$  & $6.252 \times 10^{-3}$ & $6.648 \times 10^{-3}$\\
    $\delta \omega_{440}$  & $1.047 \times 10^{-3}$ & $1.101 \times 10^{-3}$ \\
    $\delta \tau_{440}$  & $11.978 \times 10^{-3}$ &$20.28 \times 10^{-3}$ \\
    \end{tabular}
    \label{tab:FM}
\end{table}
\end{ruledtabular}

From Eq.~\eqref{eq:sigma} we see how the uncertainty varies as the inverse of the SNR, so naturally the standard deviation in the different parameters decreases as the SNR increases. Consequently, for a given source, the uncertainty is related to the total mass and the luminosity distance. To indistinctly translate luminosity distance ($D_l$) to redshift (z), we assume cosmological parameters widely used within the LISA Data Challenge (LDC) group, i.e. flat $\Lambda_{CDM}$, with Hubble constant $H_0=67.74$ and matter density $\Omega_{M} = 0.3075$, in agreement with Planck's results~\cite{Planck_2015, Planck_2020}.
We can therefore estimate the mass and the redshift needed to claim a deviation from GR with $5\sigma$. Of course, the number of sigmas $N_\sigma$ is constrained by the value of the fractional deviation itself, as shown in Eq.~\eqref{eq:sigma_dev}. 

\begin{figure*}[t]
    \centering
    \includegraphics[width=1\textwidth]{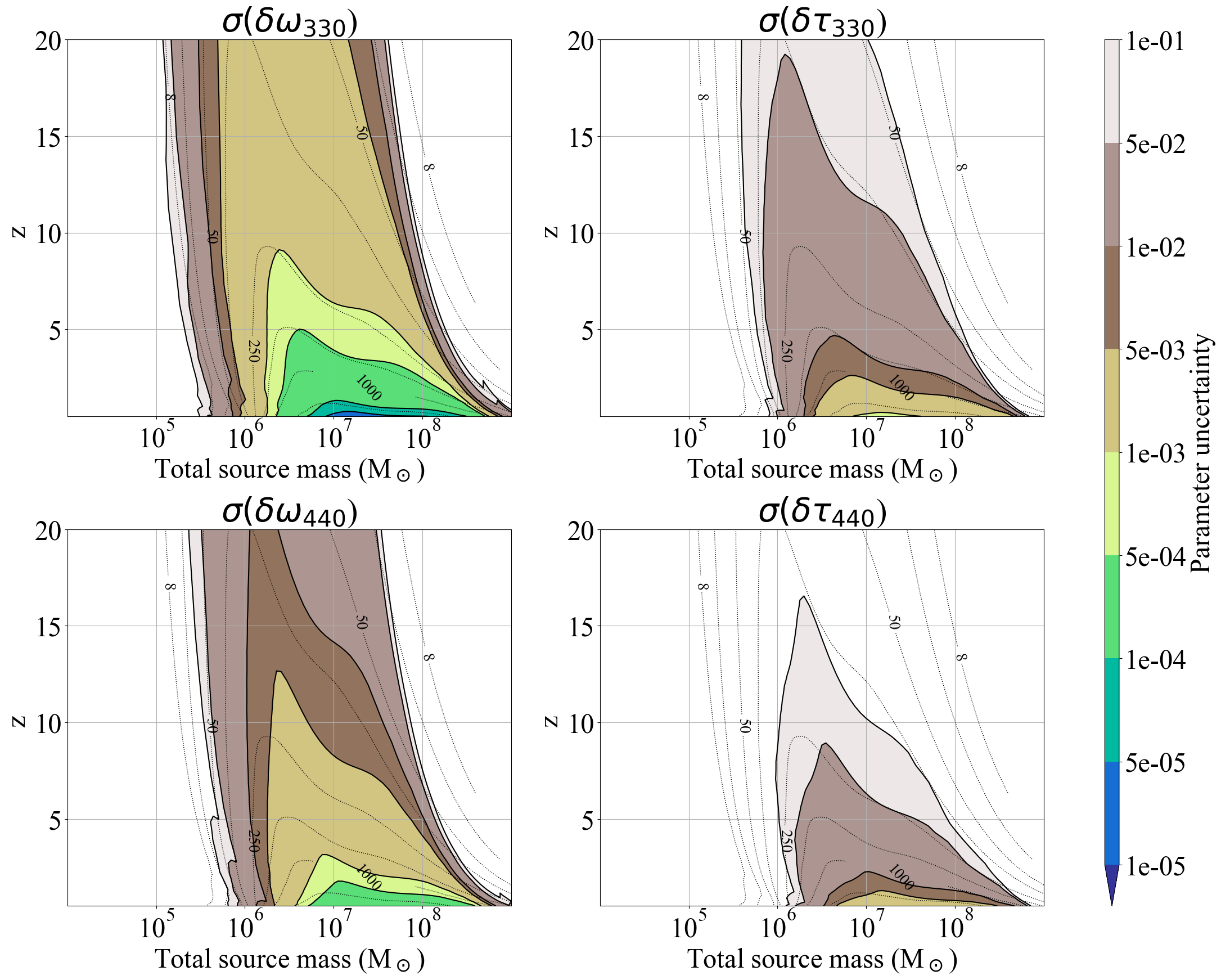}
    \caption{Uncertainty for fractional deviations from GR in $\omega$ and $\tau$ in modes (3,3,0) and (4,4,0) with respect to the source total mass and the redshift. These values are obtained for the fixed parameters listed in Table \ref{tab:params_inj}. The dotted black lines correspond to contours for a ringdown-only SNR equal to $[8, 20, 50, 100, 250, 500, 1000, 3000]$.}
    \label{fig:wtf_sigmas}
\end{figure*}

We show in Fig.~\ref{fig:wtf_sigmas} the uncertainty for the parameters with possible deviations from the GR values, such as ($\delta\omega_{330}, \delta\tau_{330}, \delta\omega_{440}, \delta\tau_{440}$), using the \textit{deviations} template. Several assumptions have been made from the beginning of the study, therefore the result we present does not provide a general detection forecast. Nevertheless, this analysis provides a qualitative understanding of LISA's ability to observe deviations from GR in the ringdown phase of a MBHB coalescence. The uncertainty on the fractional deviations is represented as a function of the source total mass and the redshift which are the dominant contributors to the SNR. We let all other source parameters fixed to the same values listed in Table \ref{tab:params_inj}. Consequently, the estimates shown in Fig.~\ref{fig:wtf_sigmas} are source-dependent, i.e., valid for the particular BH we chose as a case study. Another choice of BH parameters would change this result. Different inclinations, spins, and mass ratios would inevitably change the relative amplitude between QNMs and thus the uncertainty in each mode's complex frequency. For comparison, the value of the ringdown's SNR is also shown in the figure as contour dotted lines for the values $[8, 20, 50, 100, 250, 500, 1000, 3000]$. 

The color code on the right-hand side of Fig.~\ref{fig:wtf_sigmas} indicates the value of the uncertainty on the fractional QNM frequencies, obtained with the Fisher matrix for the considered example source. For instance, areas where $\sigma \leq 0.005$ show that LISA should be able to detect deviations from GR in $\delta\omega_{330}$ at the level of 5 standard deviations or more if the departure from GR is of the order of $0.025$ taking $\delta^{GR}= 5 \sigma$, for sources between $10^6$ to $10^7$ $M_\odot$ through the whole universe, i.e. for any possible redshift. Considering more precision-favorable situations, a deviation from GR of $5\times 10^{-5}$ would be distinguishable for sources below redshift 1 and total mass of the order of $10^7 \, M_\odot$. At first glance, one could conclude that the most severe limits will come from the deviations or lack of them in the frequencies $\delta\omega_{lmn}$ because of the high sensitivity of LISA to frequency variations. 
The expected population of MBHB for heavy seeds encloses sources in the range $[10^4 - 10^7]\, M_\odot$ up to redshift 10 approximately. This range is extended to lower-mass sources in the case of light seeds~\cite{LISA_redbook, Barausse_2020, Barausse_2021}. Hence, even if LISA cannot observe some of these \textit{golden sources}, the expectation to test the no-hair theorem and GR looks very promising for ``nearby'' sources ranging in the mass interval $[10^6 - 10^7]\, M_\odot$. 

For completeness, we performed another parameter estimation for a source with an SNR of $\sim 15$, to study whether the Fisher approximation is still valid for low SNR. The final mass and final spin were consistent with the injected values; the amplitudes and phases were consistent within 1$\sigma$ for the $(2,2,0)$ and $(3,3,0)$ modes while presenting a flat posterior for the $(4,4,0)$ mode. Regarding the fractional deviations, we recovered the $\delta \omega_{330}$ parameter with reasonable precision while this is less clear for the other deviation parameters. In Table~\ref{tab:FM_snr15}, we present the results of the estimated uncertainty for the fractional deviation and the one obtained with the Fisher matrix. When comparing them, we observe that both values partially agree for the $\delta \omega_{330}$ parameter, while it is overestimated for the other parameters. Note that the error threshold of $0.1$ avoids the mismatch for the other parameters since a larger SNR is required to achieve that precision. Nevertheless, it is important to mention that, for low SNR sources, one should be extra careful when studying the $\delta \omega_{330}$.

\begin{ruledtabular}
\begin{table}[b]
    \caption{Uncertainty computed with the Fisher matrix (second column) and obtained with the sampler (third column) for a low SNR ($\sim 15$) source.}
    \centering
    \begin{tabular}{ccc}
    & $\sigma_{\mathrm{FM}}$   & $\sigma_{\mathrm{sampler}}$  \\
    \hline
    $\delta \omega_{330}$  & $0.036$ & $0.044$\\
    $\delta \tau_{330}$  & $0.465$ & $0.127$\\
    $\delta \omega_{440}$  & $0.243$ & $0.066$ \\
    $\delta \tau_{440}$  & $1.881$ & $0.156$ \\
    \end{tabular}
    \label{tab:FM_snr15}
\end{table}
\end{ruledtabular}

\section{Conclusions}\label{sec6}

In this study, we use two approaches to explore the no-hair theorem with LISA: the \textit{agnostic} approach where no assumption on the source parameters is made except on the number of observable QNMs; and the \textit{deviations} approach, where fractional deviations of specific QNMs are estimated.

The advantage of the \textit{agnostic} approach is that it does not require any hypothesis on the events, except for the number of QNMs (which could also be inferred by performing a Bayesian model comparison not demonstrated here). We estimate the frequency and damping time for each QNM. By comparing the mass and spin derived from the complex frequencies we identify different values for each QNM, resulting in inconsistency between QNMs in the GR framework. We also quantify these deviations as fractional deviations from GR, which entails a delicate interpretation of the results depending on the assumptions made. Indeed, the non-GR-deviation hypothesis in the dominant mode is too restrictive to correctly identify the injected deviation for each QNM, despite being consistent within $2 \sigma$ with the true values. However, this kind of discrepancy can be circumvented by contrasting the results with the posterior distributions obtained from the IMR analysis, presuming that physical effects like eccentricity or others are included to avoid biased parameters. Hence this procedure requires an unbiased IMR analysis to compare results.

The \textit{deviations} approach shows better results for the fractional deviation values. However, prior assumptions are required to recover the injected values. Particularly, we assume a fixed number of observable QNMs and further constraints in the priors volume. Using the fact that no significant deviation in the first mode is observed, we do not need to rely on an IMR analysis since the intrinsic BH parameters are estimated. Including the mass and spin in the parameter estimation enables us to absorb small variations that correspond to relatively large errors in the QNMs. Thus, the hypothesis of non-deviation in the dominant mode allows us to find the injected QNM deviations confidently. If one allows for deviations in the dominant mode, extra care or further constraints in the priors are necessary due to the degeneracy between $M_f, a_f$ and $\delta\omega_{220}, \delta\tau_{220}$. Such an analysis is left for the future. 

Consequently, combining both methods could improve the characterization of possible deviations from GR. Thus, one optimized method would be to perform an \textit{agnostic} search to determine a descriptive set of QNMs and a raw estimation of the mass and spin to be compared to the IMR parameters. Once this is done, specific deviations for each QNM could be targeted, taking special care in the priors probability definition for each mode, as mode degeneracies and label switching may arise. 

Finally, we also evaluate the impact of redshift and total mass on the observable deviations from GR in the BH's spectrum with the \textit{deviations} template. From this analysis, we can estimate that in the best-case scenario, i.e., with \textit{golden sources}, the strong regime of GR could be tested up to $5 \times 10^{-3}\%$. This value is consistent with previous analysis~\cite{toubiana2024, Bhagwat_2022}. However, these sources do not dominate the estimated population of black holes in the heavy or light seeds models. Nevertheless, the prospects of testing GR in the ringdown signal from sources with masses $[10^6 - 10^7]\,M_\odot$ at redshift $\le 5$ are very promising, with a detectable fractional deviation of $1 \times 10^{-3}\% \leq \delta\omega_{330} \leq 5 \times 10^{-2}\%$ in the $(3,3,0)$ frequency's mode.


\begin{acknowledgments}

We wish to thank Sylvain Marsat for his valuable insights, and the reviewer for carefully read this article and suggest some adjustments. This work was supported by CNES, focused on the LISA Mission. We gratefully acknowledge support from the CNRS/IN2P3 Computing Center (Lyon - France) for providing computing and data-processing resources needed for this work. CP acknowledges financial support from CNES Grant No. 51/20082 and CEA Grant No. 2022-039.
\end{acknowledgments}


\bibliography{NHTL}

\end{document}